\documentclass[structabstract]{aa}
\usepackage{graphicx}
\usepackage{amssymb}
\usepackage{amsfonts}
\usepackage{txfonts}
\usepackage{natbib}
\begin{document}
\title{The star-formation history of low-mass disk galaxies: a case study of NGC\,300}

\author{X. Kang\inst{1,2,3}
  \and F. Zhang\inst{1,2}
  \and R. Chang\inst{4}
  \and L. Wang\inst{1,2,3}
  \and L. Cheng\inst{1,2,3}
         }
 \institute{Yunnan Observatories, Chinese Academy of Sciences, Kunming, 650011, China\\
 \email: {kxyysl@ynao.ac.cn}
 \and Key Laboratory for the Structure and Evolution of Celestial Objects, Chinese Academy of Sciences,
 Kunming, 650011, China
 \and University of Chinese Academy of Sciences, Beijing, 100049, China
 \and Key Laboratory for Research in Galaxies and Cosmology, Shanghai Astronomical Observatory, Chinese
 Academy of Sciences, 80 Nandan Road, Shanghai, 200030, China
 }

  \date{Received ; accepted}
\abstract
{Since NGC\,300 is a bulge-less, isolated low-mass galaxy and has not
experienced radial migration during its evolution history, it can be treated
as an ideal laboratory to test simple galactic chemical
evolution models.}
{Our main aim is to investigate the main properties of the star-formation history (SFH) of
NGC\,300 and compare its SFH with that of M33 to explore the
common properties and differences between these two low-mass systems.}
 {By assuming its disk forms gradually from continuous accretion of primordial gas
and including the gas-outflow process, we construct a simple chemical evolution model
for NGC\,300 to build a bridge between its SFH and its observed data, especially
the present-day radial profiles and global observed properties (e.g., cold
gas mass, star-formation rate and metallicity). By means of comparing
the model predictions with the corresponding observations, we adopt
the classical $\chi^{2}$ methodology to find out the best combination of
free parameters $a$, $b$ and $b_{\rm out}$. }
{Our results show that, by assuming an inside-out
formation scenario and an appropriate
outflow rate, our model reproduces well most of the present-day
observational values, not only the radial profiles but also the
global observational data for the NGC\,300 disk. Our results suggest
that NGC\,300 may experience a rapid growth of its disk.
Through comparing the best-fitting model predicted SFH of NGC\,300
with that of M33, we find that the mean stellar age of NGC\,300
is older than that of M33 and there is a lack of primordial gas
infall onto the disk of NGC\,300 recently.
Our results also imply that the local environment may paly a key
role in the secular evolution of NGC\,300.}
{}
 \keywords{galaxies: abundances --- galaxies: evolution--- galaxies: spiral
--- galaxies: individual: NGC\,300 }

\titlerunning{The star-formation history of NGC\,300}
\authorrunning{X. Kang et al.}
\maketitle
%
\section{Introduction}
\label{sec_intro}
In order to fully understand the history of the universe, it is
essential to understand the history of individual galaxies.
However, most of disk galaxies are difficult to study in detail
due to the complexities of their bulges \citep{B&F95}, or
tidal interactions, or mergers \citep{B&H92}. Fortunately,
there still exist some disk galaxies, which are in the absence
of major mergers, undisturbed and
bulge-less (i.e., pure-disk) systems that maintain weak spiral
structure to suppress radial migration, making them excellent
laboratories to study the secular evolution of galactic
disk. At the same time, recent studies have exhibited that
massive galaxies have almost finished forming stars at early
epoch and a large number of low-mass galaxies exist at all redshifts
\citep[e.g.,][and references therein]{bauer13}. Moreover, deep
surveys detecting masses $M_{\ast}<10^{10}\,{\rm M_{\odot}}$
have shown that galaxy stellar mass functions have steep
slopes at the low-mass end \citep{baldry08,baldry12,kelvin14},
indicating the necessity to understand the evolutionary history
of low-mass galaxies, which are the most populous in the universe.

NGC\,300 is the nearest isolated and nearly
face-on low-mass disk galaxy,
which is located at a distance of 2.0\,Mpc \citep{Dalcanton09} in the
Sculptor group. In addition, NGC\,300 is an almost bulge-less system
\citep{williams13a}, and there is no confusion between
disk and bulge populations \citep[bulge light $<$ 20\%,][]{vlajic09}.
Both N-body simulations
and the kinematics of the globular clusters systems studies have shown
that NGC\,300 has not experienced radial migrations during its evolution
\citep{olsen04,gogarten10,nantais10}. Therefore, NGC\,300 is an
ideal disk galaxy to test the simple galactic chemical evolution
model.

The simple chemical evolution model, which adopts parameterized
descriptions for some complicated physical processes, has already been
proven to be a powerful tool to explore the formation
and evolution of disk galaxies
\citep{Tinsley80,chang99,bp00,chiappini01,m&d05,yin09,chang12,kang12,kang15},
and it has achieved great progress in our understanding
the formation and evolution of disk galaxies in a cosmological context.
However, there is still lack of such research on NGC\,300.
In this paper, we construct a simple chemical evolution model for
NGC\,300 to build a bridge between its SFH and its observed properties,
including both the radial profiles and global observational
constraints.

More importantly, NGC\,300 is also a near-optical twin of
the Local Group galaxy M33, and Table \ref{Tab:obs1} presents
a comparison of their basic properties. Several authors
have found an inside-out growth scenario for NGC\,300 based on
its broadband colors and the color-magnitude diagram (CMD)
fitting \citep{kim04,MM07,gogarten10}. Both theoretical model
\citep{kang12} and
observations \citep{Williams09} have shown that M33 grows
inside-out. Although they are similar in appearance,
M33 has a disk break at $\sim8\,\rm kpc$
\citep{Ferguson07,barker07,barker11}, while NGC\,300 has
a pure exponential disk out to $\sim14\rm kpc$ \citep{bland05}.
Furthermore, an HI bridge exists between M33 and M31
\citep{karachentsev04,putman09}, indicating the probability that they
interacted with each other in the past \citep{braun04,McConnachie10,sanroman10,Bernard12}.
This result is confirmed by the evidence for tidal disruption of M33's
gas disk \citep{Rogstad76,Deul87,Corbelli97,karachentsev04,putman09}.
Compared to M33, NGC\,300 is a relatively isolated system, with
only dwarf galaxies nearby \citep{karachentsev03,Tully06,williams13a}.
The aforementioned differences indicate that they may have
significantly different evolution histories, and it should be
interesting to compare the SFH of NGC\,300 with that of
M33.

The outline of this paper is organized as follows. The
observational constraints are presented in Section 2.
The main ingredients of our model are described in Section 3.
Comparisons of the model predictions with the present-day
observations of NGC\,300, as well as comparisons between the
SFH of NGC\,300 and that of M33 are shown in Section 4.
Section 5 presents a summary of our results.

\begin{table}
\caption{Basic properties of NGC\,300 and M33.}
\label{Tab:obs1}
\begin{center}
\begin{tabular}{lll}
\hline
\hline
Property             &   NGC\,300                 &   M33\\
\hline
Morphology               &  SA(s)d$^{\rm a, b}$              & SA(s)cd$^{\rm a, b}$  \\
Distance                 &  2.0\,Mpc$^{\rm c}$            & 800\,kpc$^{\rm d}$  \\
$M_{\rm B}$              &  $-17.66^{\rm c}$              & $-18.4^{\rm e}$  \\
$M_{\rm K}$              &  $-20.1^{\rm f}$               & $-20.4^{\rm f}$  \\
Scale-length\,(kpc, in $K-$band)      &  1.29$^{\rm g}$                 &  1.4$^{\rm g}$   \\
Rotation velocity        &  91\,km\,s$^{-1\rm \,h}$         &  110\,km\,s$^{-1\rm \,h}$  \\
\hline
\end{tabular}\\
\end{center}
Refs: (a) NED; (b) \citet{1991S&T....82Q.621D};
(c) \citet{Dalcanton09};
(d) \citet{Williams09}; (e) \citet{costas92}; (f) \citet{jarrett03};
(g) \citet{MM07}; (h) \citet{garnett02}
\end{table}

\section{Observational constraints}
\label{sect:Obs}

A successful chemical evolution model of NGC\,300, especially
one involving free parameters, should reproduce as many as
possible observational constraints, including both local
(concerning the radial profiles) and global ones
(concerning the whole disk).
It is well known that, during the evolutionary histories of
galaxies, star formation process converts
cold gas into stars and stars in turn produce heavy elements
by means of nucleosynthesis. These metals are then expelled into
the surrounding ISM in the later stage of stellar evolution,
thus enriching interstellar gas and becoming fuel for the next
generation of star formation.
That is, the observed present-day metallicity, atomic hydrogen
gas and star-formation rate (SFR) provide important
constraints on the SFH of the galaxy.

\subsection{Cold gas and SFR}

The atomic hydrogen gas disk associated with NGC\,300 is
observed to extend far beyond its optical disk.
The radial distribution of atomic
hydrogen gas for NGC\,300 disk is measured with the Australia
Telescope Compact Array (ATCA) at a wavelength
of $\lambda\,=\,21\,\rm cm$ \citep{westmeier11}, and its atomic
hydrogen mass has been estimated to be
$M_{\rm HI}\sim(1.1-1.87)\times10^{9}\rm M_{\odot}$
\citep{rogstad79,barnes01,Koribalski04,westmeier11,millar11,Wiegert14}.
The stellar mass of the NGC\,300 disk is estimated to be
$M_{\rm \ast}\sim 1.928\times10^{9}\rm M_{\odot}$ \citep{MM07}.
Thus, we can easily obtain the atomic hydrogen gas fraction (defined as
$f_{\rm gas}\,=\,\frac{M_{\rm H_I}}{M_{\rm H_I}+M_{\rm *}}$)
of NGC\,300, i.e., $\sim0.363-0.492$.

The recent SFR surface density profiles of \citet{gogarten10}
are obtained from
the resolved stars, and those from \citet{williams13a} are derived
by the combination of far-ultraviolet (FUV) and  $24\,\mu\rm m$ maps.
In recent years, the total SFR in several regions along the NGC\,300
disk have been carefully measured by both ground-based and space-based
facilities \citep{helou04,gogarten10,binder12,K&K13,williams13a}.
The current total SFR for NGC\,300 disk is estimated to be
$0.08-0.30\rm M_{\odot }$\,yr$^{-1}$ using different tracers,
including the X-ray luminosity \citep{binder12}, H${\alpha}$
emission \citep{helou04,K&K13}, and luminosity in the FUV
\citep{K&K13} and mid-infrared \citep[MIR,][]{helou04}
bands, just a factor of several smaller than the value in
the Milky Way disk, which indicates that NGC\,300 is currently
a rather quiescent galaxy.

\begin{table}
\caption{Global observational constraints for the disk of NGC\,300.}
\label{Tab:obs2}
\begin{center}
\begin{tabular}{lll}
\hline
\hline
Property             &      Value                &    Refs.      \\
\hline
Stellar mass         & $\sim1.928\times10^{9}~\rm M_{\odot}$         & 1\\
H{\sc i} mass        & $\sim(1.10-1.87)\times10^{9}~\rm M_{\odot}$    & 2, 3, 4, 5, 6, 7 \\
Gas fraction         & $\sim0.363-0.492$                                   & this paper \\
${\rm log(O/H)+12}$  & $\sim8.40-8.56$                                   & 8, 9, 10 \\
Total SFR            & $\sim0.08-0.3~\rm M_{\odot }$\,yr$^{-1}$              & 11, 12, 13, 14 \\
\hline
\end{tabular}\\
\end{center}
Refs:
(1) \citet{MM07}; (2) \citet{rogstad79}; (3) \citet{barnes01};
(4) \citet{westmeier11}; (5) \citet{Wiegert14};
(6) \citet{Koribalski04}; (7) \citet{millar11}; (8) \citet{garnett02};
(9) \citet{bresolin09}; (10) \citet{pilyugin14_1}; (11) \citet{helou04};
(12) \citet{gogarten10} (13) \citet{binder12}; (14) \citet{K&K13}
\end{table}

\subsection{Chemical abundance}

Observations of the chemical
abundances of a galaxy provide important constraints for the physical
processes in the models of galaxy evolution, since the chemical
composition is directly related to star formation process. In this case,
planetary nebulae (PNe) and H{\sc ii} regions are good tracers.
Generally speaking, they are representative of the chemical
abundance in the interstellar medium (ISM) at two different
epochs during the galaxy
evolution. PNe proceed from the death of low- and intermediate-mass stars
($1{\rm M_{\odot}}\,<\,M\,<\,8{\rm M_{\odot}}$), so they have oxygen
abundances approximately equal to the composition of the ISM at the epoch when
their progenitors were formed. On the other hand, H{\sc ii} regions
are thought to represent the current chemical composition of the
ISM. Moreover, oxygen is the most abundant heavy element formed
in the Universe \citep{korotin14,zahid14}, and oxygen abundance
is easily estimated in H{\sc ii} regions due to its bright
emission line. Therefore, the abundance of oxygen can be
used as a proxy for the production of all heavy elements in
galaxies \citep{zahid14}.

Radial gas-phase oxygen profiles are regarded as important
quantities to constrain our model. The radial gas-phase oxygen
abundance for NGC\,300 in the H{\sc ii} regions and PNe
has been studied by a number of authors \citep{pagel79,Deharveng88,
zaritsky94,bresolin09,stasinska13,pilyugin14_1}.
All these studies showed that oxygen abundance decreases with radius,
and the metallicity in the center is highest, although the overall
abundance depends on the calibration methods adopted. The observed
oxygen abundance profiles of the NGC\,300 disk used
to constrain the gas-phase oxygen predicted by our model are
those of \citet{bresolin09} and \citet{pilyugin14_1}, since
these two papers enlarged the number of the H{\sc ii} regions
in whose spectra [O{\sc iii}]$\lambda4363$ can be detected.

At the same time, the characteristic oxygen abundance, which is
defined as the oxygen value $\rm 12+log(O/H)$ at the effective radius
$r_{\rm eff}$, is representative of the average value
of the young component across the galaxy \citep{garnett02,sanchez13}.
$r_{\rm eff}$ is equal to 1.685 times the exponential scalelength
$r_{\rm d}$ of the disk. Thus, we will also use the characteristic oxygen
abundance of \citet{bresolin09} and \citet{pilyugin14_1} to compare
with the model predicted present-day oxygen
abundance of NGC\,300 to constrain our chemical evolution model.

In Table \ref{Tab:obs2}, we present the basic observational constraints
concerning the present-day total amount of stars, atomic hydrogen gas,
gas fraction, characteristic oxygen abundance and SFR in the disk
of NGC\,300.

\section{The Model}
\label{sect:analysis}

In this section, we briefly introduce the basic assumptions and main
ingredients of the model. We assume that the NGC\,300 disk is composed
of a set of independently concentric rings with the width 500
pc, which are progressively built up by continuous infall of primordial gas
($X\,=\,0.7571, Y_{\rm p}\,=\,0.2429, Z\,=\,0$) from its halo.
Star formation, metal production via stellar evolution, stellar mass return,
infalls of primordial gas, and outflows of metal enriched gas are taken into
account in our model under the condition of both instantaneous recycling
assumption (IRA) and instantaneous mixing of the ISM with stellar ejecta.
Moreover, neither radial gas flows nor stellar migration is allowed
in our model, since these radial flows are far from clearly understanding
and will introduce additional free parameters whose values might be
difficult to constrain.

\subsection{Gas infall rate}

In order to reproduce the observed metallicity distribution of
long-living stars in the solar neighborhood
\citep[G-dwarf problem,][and references therein]{pagel89,rana91},
the simple closed-box model is proved to be unreasonable and
the infall of gas from its halo has been introduced in the
chemical evolution model \citep{Tinsley80}. Indeed, \citet{mathewson75}
conclude that a long 'H{\sc i} tail' extends about $2^{\circ}$ to the
southeast of the disk of NGC\,300 and \citet{westmeier11} find that
the H{\sc i} disk is more extended than its optical disk, respectively.

The infall rate, as a function of space and time, plays a key role
in determining the properties of the galaxy disk, such as the gas,
SFR and metallicity. At a given radius $r$, the gas infall rate
$f_{\rm in}(r,t)$ (in units of $\rm{M_{\odot}}\,{pc}^{-2}\,{Gyr}^{-1}$)
is assumed to be \citep{kang12}:
\begin{equation}
f_{\rm{in}}(r,t)=A(r)\cdot t\cdot e^{-t/\tau},
\label{eq:infall rate}
\end{equation}
where $\tau$ is the infall time-scale and is a free parameter
in our model. The $A(r)$ are actually a set of separate quantities
constrained by the present-day stellar mass surface density
$\Sigma_*(r,t_{\rm g})$, and $t_{\rm g}$ is the cosmic age
and we set $t_{\rm g}=13.5\rm\,Gyr$ according to the standard
cosmology (e.g., $H_{0}\,=\,70\,\rm{km\,{s}^{-1}\,{Mpc}^{-1}}$,
$\Omega_{\rm M}\,=\,0.3$, and $\Omega_\Lambda\,=\,0.7$).
In other words, $A(r)$ are iteratively computed by requiring the
model predicted present-day stellar mass surface density
is equivalent to its observed value \citep{chang12,kang12}.
In general, the $K$-band luminosity strongly correlates with the
accumulated star formation in the galaxy, thus it can accurately
reflect the stellar disk. In this work, the present-day values of
stellar mass and disk scale-length for NGC\,300 are adopted to be
$M_{*}=1.928\times10^{9}{\rm M_{\odot}}$ and $r_{\rm d}=1.29$\,kpc,
and these two values are derived from $K-$band luminosity
\citep{MM07}. We assume that NGC\,300 has a pure
exponential disk, which is confirmed by the observations of \citet{bland05},
and $\Sigma_*(r,t_{\rm g})$ follows an exponential profile, thus
the central stellar mass surface density can be easily obtained from
$\Sigma_*(0,t_{\rm g})=M_{*}/2\pi r_{\rm d}^{2}$.

\begin{figure}
  \centering
  \includegraphics[angle=0,scale=0.475]{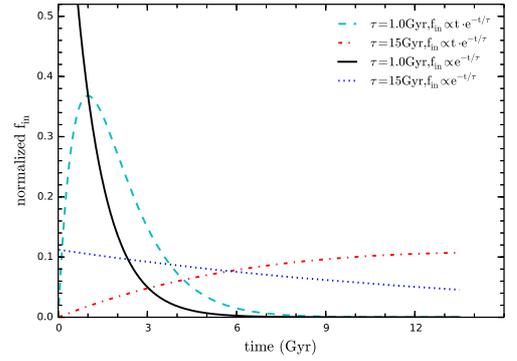}
    \caption{The normalized gas infall rate with the two forms.
    The infall rate adopted as $f_{\rm{in}}(r,t)\propto t\cdot e^{-t/\tau}$
    is plotted as dashed line ($\tau=1.0\,{\rm Gyr}$) and dot-dashed line
    ($\tau=15\,{\rm Gyr}$), while that adopted as
    $f_{\rm{in}}(r,t)\propto e^{-t/\tau}$ is shown as solid line ($\tau=1.0\,{\rm Gyr}$)
    and dotted line ($\tau=15\,{\rm Gyr}$), respectively.
    }
  \label{Fig:normalized}
\end{figure}

It should be emphasized that a simple exponential form of gas
infall rate is widely used in the previous chemical evolution models
\citep{matteucci89,hou00,chiappini01,yin09}. To explore the
difference between the exponential gas infall rate form and
the form adopted in this paper, we show the two forms of the
normalized gas infall rate in Fig. \ref{Fig:normalized}.
The infall rate
adopted as $f_{\rm{in}}(r,t)\propto t\cdot e^{-t/\tau}$ is
plotted as dashed line ($\tau=1.0\,{\rm Gyr}$) and dot-dashed line
($\tau=15\,{\rm Gyr}$), while that adopted as $f_{\rm{in}}(r,t)\propto e^{-t/\tau}$
is shown as solid line ($\tau=1.0\,{\rm Gyr}$)
and dotted line ($\tau=15\,{\rm Gyr}$), respectively. From
Fig. \ref{Fig:normalized}, we can see that the gas infall
rate we adopted in this work is low in the beginning and
gradually increases with time, it reaches the peak value
when $t=\tau$ and then falls, that is, $\tau\,\rightarrow\,0$ is
corresponding to time-declining gas infall rate, while
$\tau\,\rightarrow\,\infty$ is corresponding to a time-increasing
gas infall rate. However, the exponential
form of gas infall rate is always decreasing with time
and the limiting case is corresponding to a constant.
In order to include more possible situations in the model,
we adopt the gas infall
rate as the form of $f_{\rm{in}}(r,t)\propto t\cdot e^{-t/\tau}$
in this work.
In any case, the infall timescale $\tau$ determines the gas
accretion history and largely describes the main properties
of the SFH of the disk of NGC\,300, thus we regard $\tau$ as
the most important free parameter in our model.

\subsection{Star-formation Law}

The star-formation law, which connects SFR surface density with
interstellar gas surface density, is another significant
ingredients of the model, and SFR describes the total mass
of newly born stars in unit time.

More than half a century ago, \citet{schmidt1959} surmised that the
SFR surface density should vary with a power $n$ of the surface
density of gas. Later,
based on the observed data of a sample of 97 nearby normal
and starburst galaxies, \citet{k98} demonstrated the existence
of a power-law between the galaxy-averaged SFR surface
density and the galaxy-averaged total gas surface density.
Recently, high-resolution investigations that separate the components of
atomic hydrogen and molecular hydrogen gas have shown that the SFR
surface density correlates more strongly with the surface density of
molecular hydrogen than with that of atomic hydrogen and total
gas \citep{wong02,bigiel08,leroy08}.
Indeed, multiple studies in both individual galaxies and large
samples of spiral galaxies have found a linear relation between
the SFR surface density $\Psi(r,t)$ and the molecular gas surface density
$\Sigma_{\rm H_2}(r,t)$. The linear slope is an approximately constant
depletion time
\citep[$t_{\rm dep}=\Sigma_{\rm H_2}(r,t)/\Psi(r,t)$, e.g.,][]{bigiel08,leroy08,rahman11,leroy13}.
Furthermore, based on an up-to-date set of observational data,
\citet{krumholz14} concludes in his recent work that
"the correlation between star formation and molecular gas is
the fundamental one". These results do encourage us to adopt
the molecular related star formation law here , that is
\begin{equation}
\Psi(r,t)=\Sigma_{\rm{H_2}}(r,t)/t_{\rm dep},
\label{eq:h2sfr}
\end{equation}
With respect to the value of molecular gas depletion time $t_{\rm dep}$,
we adopt $t_{\rm dep}\,=\,1.9\,\rm Gyr$ throughout this work
\citep{leroy08,leroy13}.
The reader is referred to \citet{kang12} and \citet{kubryk14} for a
more indepth description for the calculation of the ratio of molecular
hydrogen to atomic hydrogen gas surface density $R_{\rm mol}$ in a galaxy
disk, along with the present-day molecular hydrogen and
atomic hydrogen gas surface density.

\subsection{Gas outflow rate}

The gas-outflow process may influence the evolutionary history of
NGC\,300. The oxygen in the universe is predominantly formed in massive
stars ($>8\,\rm{M_{\odot}}$) and subsequently dispersed into the
ISM by supernova explosion and stellar winds,
thus the oxygen in the ISM may expel from the galactic disk when the
gas thermal energy exceeds the binding energy of gas. Low-mass
galaxies are expected to lose a larger fraction of supernova
ejecta than high-mass systems due to their shallower gravitational
potentials \citep{kauffmann93}. Indeed, both analytical \citep{Erb08}
and hydrodynamic \citep{Finlator08} models showed that galactic
outflows are important to reproduce the stellar mass-metallicity
relation of galaxies. Moreover, for galaxies with stellar mass
$M_{*}\leq\,10^{10.5}\,\rm {M_{\odot}}$, the outflow process play
a crucial role during their evolution histories
\citep{tremonti04,spitoni10,chang10}.
Finally, galaxies with rotation speed $V_{\rm rot}\leq\,100-150\,{\rm km s^{-1}}$
may expel a large part of their supernova ejecta to the
circumgalactic medium \citep{garnett02}. Since
NGC\,300 is a fairly low-mass disk galaxy
($M_{\ast}\approx\,10^{9.29}\,\rm {M_{\odot}}$, \citet{MM07})
with a rotation speed about $V_{\rm rot}\approx 91\,{\rm km s^{-1}}$
\citep{garnett02}, the gas-outflow process has a significant influence on the
chemical enrichment during its evolution history.

We assume that the outflowing gas has the same metallicity
as the ISM at the time the outflow is launched, and the
outflowing gas will not fall again to the disk
\citep{chang10,kang12,ho15}. We follow the approach of
\citet{recchi08}, that is, the gas outflow rate
$f_{\rm out}(r,t)$ (in units of $\rm{M_{\odot}}\,{pc}^{-2}\,{Gyr}^{-1}$)
is proportional to the SFR surface density $\Psi(r,t)$. Therefore,
the outflow rate is given by:
\begin{equation}
  f_{\rm out}(r,t)=b_{\rm out}\Psi(r,t)
\label{eq:outflow}
\end{equation}
where $b_{\rm out}$ is the other free parameter in our model.

\subsection{Basic equations of chemical evolution}

As mentioned above, both IRA and instantaneous mixing
of the ISM with ejecta are assumed, thus the chemical evolution
in each ring can be expressed by the classical set of
integro-differential equations from \citet{Tinsley80}:

\begin{equation}
\frac{{\rm d}[\Sigma_{\rm tot}(r,t)]}{{\rm d}t}\,=\,f_{\rm{in}}(r,t)-f_{\rm{out}}(r,t),\\
\label{eq:tot}
\end{equation}
\begin{equation}
\frac{{\rm d}[\Sigma_{\rm gas}(r,t)]}{{\rm d}t}\,=\,-(1-R)\Psi(r,t)+f_{\rm{in}}(r,t)-f_{\rm{out}}(r,t),\\
\label{eq:gas}
\end{equation}
\begin{eqnarray}
\frac{{\rm d}[Z(r,t)\Sigma_{\rm gas}(r,t)]}{{\rm d}t}\,=\,y(1-R)\Psi(r,t)-Z(r,t)(1-R)\Psi(r,t) \nonumber\\
+Z_{\rm{in}}f_{\rm{in}}(r,t)-Z_{\rm{out}}(r,t)f_{\rm{out}}(r,t).
\label{eq:metallicity}
\end{eqnarray}
where $\Sigma_{\rm tot}(r,t)$ is the total (star + gas) mass surface
density in the ring centered at galactocentric distance $r$ at
evolution time $t$, and $Z(r,t)$ is the metallicity in the corresponding
place and time. $R$ is the return
fraction and we get $R=0.3$ after adopting stellar initial mass
function (IMF) from \citet{Chabrier03} between $0.1\,{\rm M}_{\odot}$
and $100\,{\rm M}_{\odot}$. $y$ is the nucleosynthesis yield and we set
$y=1\,{\rm Z}_{\odot}$ throughout this work \citep{fu09,chang10,kang12}.
$Z_{\rm{in}}$ is the metallicity
of the infalling gas and we assume the infalling gas is metal-free,
i.e., $Z_{\rm{in}}=0$. $Z_{\rm{out}}(r,t)$ is the metallicity
of the outflowing gas and we assume that the outflow gas
has the same metallicity as the ISM at the time the outflows are launched,
e.g., $Z_{\rm{out}}(r,t)=Z(r,t)$ \citep{chang10,ho15}.

In summary, the infall time-scale $\tau$ and the outflow
coefficient $b_{\rm out}$ are two free parameters in our model.
Moreover, there is degeneracy between $y$ and the outflow parameter
$b_{\rm out}$ in that the model adopting a higher $y$ need a
larger $b_{\rm out}$ to reproduce the observed metallicity profile.
Thanks to the fact that the reasonable range of y is small
(about $0.5\,{\rm Z}_{\odot}$ to $1\,{\rm Z}_{\odot}$) compared with the
large possible rang of $b_{\rm out}$, we can constrain $b_{\rm out}$ using
the observed abundance gradient. We should emphasize that, although
the accurate value of free parameters in our best-fitting model
may change a little, our results of the main trends of the
SFHs of NGC\,300 are robust.

\begin{figure*}
   \centering
   \includegraphics[angle=0,scale=0.9]{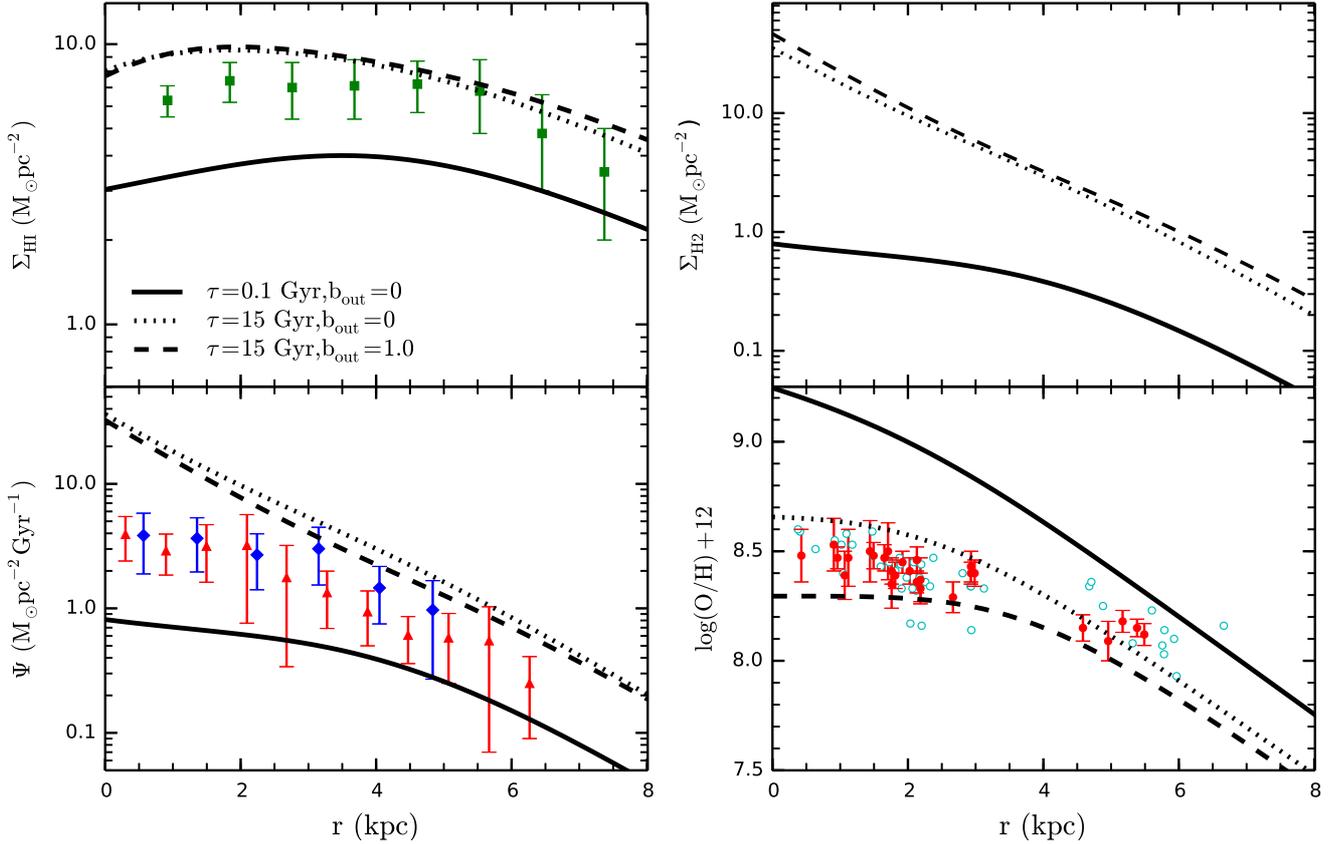}
   \caption{The influence of infall time-scale $\tau$ and outflow
   efficiency $b_{\rm out}$ on the model
   results. Different line types are corresponding to various parameter
   groups: solid lines $(\tau, b_{\rm out})=(0.1\,{\rm Gyr}, 0)$,
   dotted lines $(\tau, b_{\rm out})=(15\,{\rm Gyr}, 0)$, dashed lines
   $(\tau, b_{\rm out})=(15\,{\rm Gyr}, 1)$.
   On the left-hand side, the  radial profiles of H{\sc i} and SFR
   surface density are separately displayed in the top and bottom panels;
   On the right-hand side, the radial profiles of H2 and oxygen abundance
   are shown in the top and bottom panels, respectively.
   H{\sc i} data from \citet{westmeier11} are shown by
   filled squares. SFR data taken from \citet{williams13a} are denoted
   by filled triangles and those taken from \citet{gogarten10} are
   shown by filled diamonds. Note that the SFR data from
   Gogarten et al. (2010) are the recent SFR as a function of radius.
   The observed oxygen abundance from \citet{pilyugin14_1}
   are shown as open circles, while those from \citet{bresolin09} are plotted
   by filled circles.
   }
   \label{Fig:results1}
   \end{figure*}

\section{Model Results Versus Observations}
\label{sect:result}

In this section, we present our results step by step. Firstly,
we investigate the influence of free parameters on model predictions
and select a best-fitting model for NGC\,300. Then, we present the
evolution of the total amount of atomic
hydrogen gas, SFR, characteristic $\rm 12+log(O/H)$, and of the
gas fraction for the disk of NGC\,300. Finally, we compare the
SFH of NGC\,300 with that of M33.

\subsection{Radial profiles}

First of all, we explore the influence of infall time-scale $\tau$ on
model results and fix the gas-outflow coefficient to be $b_{\rm{out}}=0$.
The comparison between the profiles of model predictions and the
observations are displayed in Fig. \ref{Fig:results1}.
The different line types in Fig. \ref{Fig:results1} are
corresponding to various values of free parameters: solid lines
($\tau=0.1\,{\rm Gyr}$) and dotted lines ($\tau=15.0\,{\rm Gyr}$).
Obviously, the model predictions are
very sensitive to the adopted infall time-scale $\tau$. As shown
in Fig. \ref{Fig:normalized}, the case of $\tau=0.1\,{\rm Gyr}$ corresponds
to a time-declining infall rate that most of the cold gas has been
accreted to the disk in the early stage of of its history, while that
of $\tau=15.0\,{\rm Gyr}$ denotes a time-increasing gas infall rate
that a large fraction of cold gas is still falling onto the disk of NGC\,300 nowadays.
Thus, the model adopting a shorter infall time-scale (solid lines)
predicts lower gas surface density, lower SFR, and
higher gas-phase oxygen abundance than that adopting a longer
infall time-scale (dot-dashed lines). This is mainly due to the fact
that, in our model, the setting of longer infall time-scale corresponds
to a slower gas accretion during its evolution history, thus more
gas remain in the disk nowadays to fuel higher SFR and then
results in an younger stellar population, lower metallicity and
higher cold gas content at the present time, and vice versa.

Here, we investigate the impact of the gas-outflow
process on model predictions. The dashed lines in Fig.
\ref{Fig:results1} represent the model results adopting
$(\tau, b_{\rm out})=(15.0\,{\rm Gyr}, 1.0)$. The comparison between
the dotted ($\tau\,=15.0\,{\rm Gyr}, b_{\rm out}=0$)
and dashed ($\tau=15.0\,{\rm Gyr}, b_{\rm out}=1.0$)
lines shows that the gas-outflow process has little influence
on H{\sc i}, H$_{2}$ and SFR, while it has great influence on
the gas-phase metallicity. That is, the gas-outflow process carries
part of metals away from the disk and reduces the gas-phase metallicity.
In other words, the observational radial gas-phase metallicity must be very
important physical quantities to constrain the gas-outflow
process of NGC\,300.

Fig. \ref{Fig:results1} also shows that the area between
the solid ($\tau\,=\,0.1\,{\rm Gyr}, b_{\rm out}=0$) and dashed
($\tau\,=\,15.0\,{\rm Gyr}, b_{\rm out}=1.0$) lines almost covers
the whole range of the observations, which suggests the possibility to
construct a model that can reproduce most of the observed
features along the disk of NGC\,300. Furthermore, the observed
trends in Fig. \ref{Fig:results1} show that the inner stellar
disk is metal-richer than that of outer region, and the observed
studies also show that the stellar disk of NGC\,300 grows inside-out
\citep{kim04,MM07,gogarten10}. To simply describe the inside-out formation
scenario of disk, we follow the previous models of \citet{chiosi1980}
and \citet{matteucci89}, and adopt a radius-dependent infall time-scale
$\tau(r)\,=\,a\times r/{\rm r_{d}}+b$,
where a and b are the coefficients for the linear equations adopted
for $\tau(r)$. Including the additional free parameter $b_{\rm out}$
in the gas-outflow process, there are three free parameters (a, b and
$b_{\rm out}$) in our model.


   \begin{figure}
  \centering
  \includegraphics[angle=0,scale=0.475]{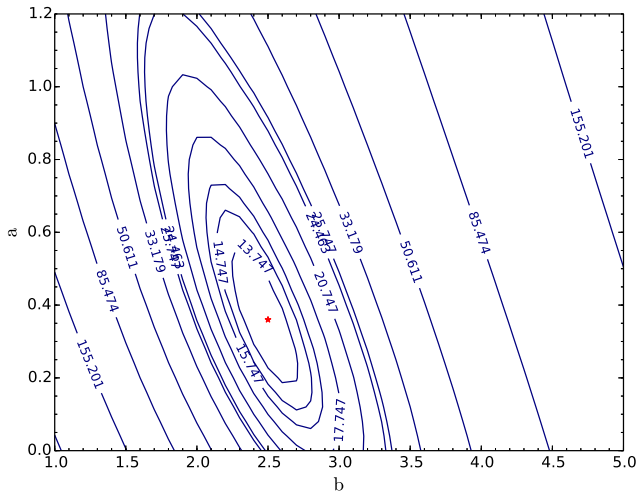}
    \caption{$\chi^{2}$ contours plot of a, b and
    $b_{\rm out}$ determination. $\chi^{2}$ contours are calculated
    by allowing $b_{\rm out}$ to vary to minimize $\chi^{2}$ for each pair
    of values of a and b. The number of each contour shows
    the value of $\chi^{2}$. The first, second and third ellipses are
    contours that correspond to $68\%$, $84\%$ and $92\%$ confidence
    level, respectively.
    }
  \label{Fig:contour}
\end{figure}

\begin{figure*}
   \centering
   \includegraphics[angle=0,scale=0.9]{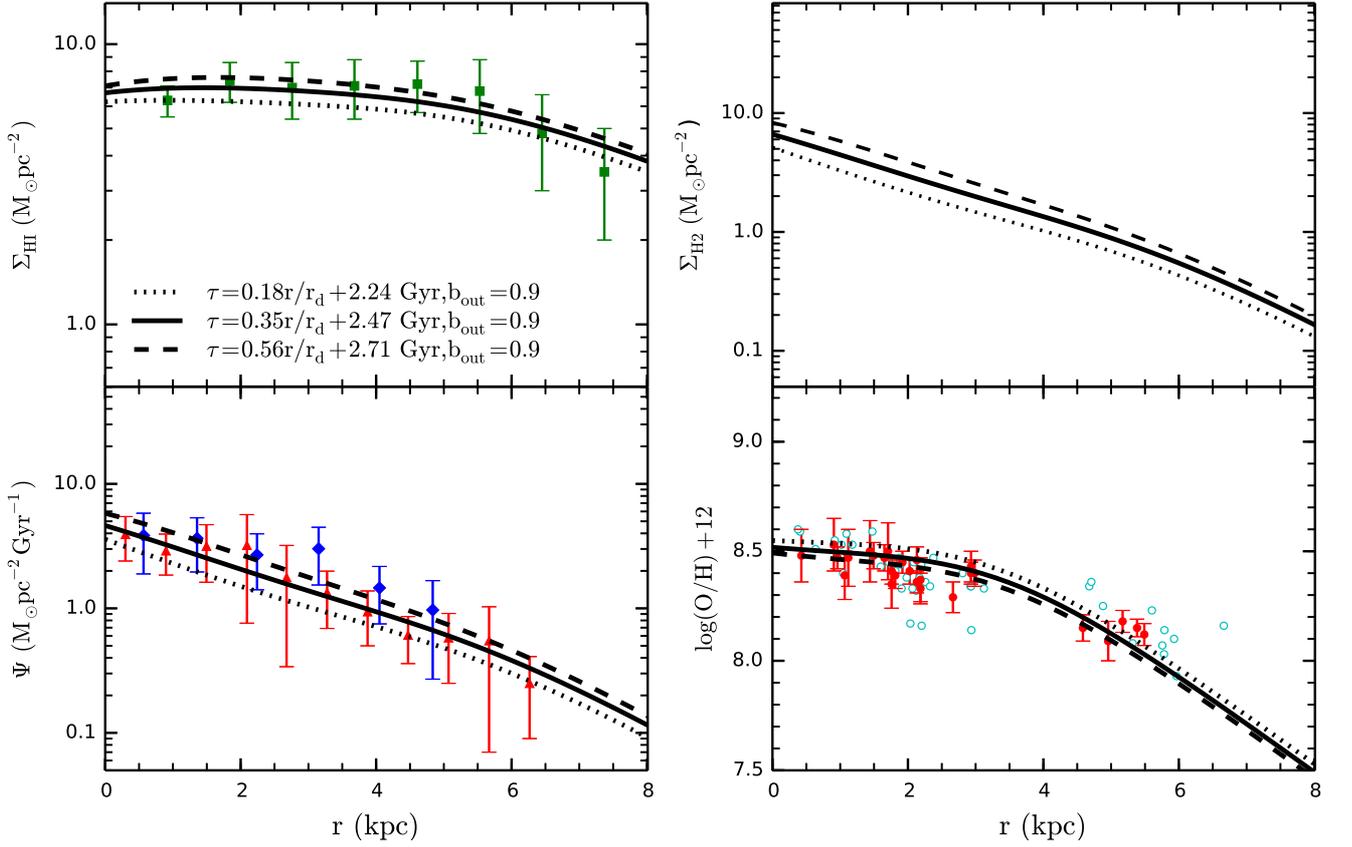}
   \caption{Comparisons of the model predictions with the
   observations. The solid lines represent the best-fitting model results
   adopting $(\tau, b_{\rm out})=(0.35r/{\rm r_{d}}+2.47\,{\rm Gyr}, 0.9)$,
   while the dotted and dashed lines plot the model predictions adopting
   $(\tau, b_{\rm out})=(0.18r/{\rm r_{d}}+2.24\,{\rm Gyr}, 0.9)$ and
   $(\tau, b_{\rm out})=(0.56r/{\rm r_{d}}+2.71\,{\rm Gyr}, 0.9)$
   with $68\%$ confidence level, respectively. The notation of the observational
   data is the same as that of Fig. \ref{Fig:results1}. }
   \label{Fig:results1_1}
\end{figure*}

In order to find the best combination of free parameters a, b and
$b_{\rm out}$, we adopt the classical $\chi^{2}$ technique to
compare the model results with the corresponding observational
data, including the radial profiles of H{\sc i}, SFR and
$\rm 12+log(O/H)$. The boundary conditions of a, b and $b_{\rm out}$
are assumed to be $0\leq a\leq1.2$, $1.0\leq b\leq5.0$ and
$0<b_{\rm out}\leq0.9$, respectively. In practice, we calculate the
value of $\chi^{2}$ by comparing our model predictions with the observed data
(i.e., the combination of the radial
profiles of H{\sc i}, SFR and $\rm 12+log(O/H)$), and show the $\chi^{2}$
contours in Fig. \ref{Fig:contour}, where the number of each contour shows
the value of $\chi^{2}$.
The first, second and third ellipses are contours that correspond
to $68\%$, $84\%$ and $92\%$ confidence level, respectively.
It should be emphasized that,
for each pair of a and b values, we change the value of $b_{\rm out}$ in
order to make sure the value of $\chi^{2}$ minimum. The minimum value of
$\chi^{2}$ (12.747) denoted as filled asterisk is displayed in Fig.
\ref{Fig:contour}, and the corresponding values of the best combination
(a, b, $b_{\rm out}$)= (0.35, 2.47 0.9) is selected as the best-fitting model.

The predictions of best-fitting model are shown as solid
lines in Fig. \ref{Fig:results1_1}.
The dotted and dashed lines plot the model predictions adopting
$(\tau, b_{\rm out})=(0.18r/{\rm r_{d}}+2.24\,{\rm Gyr}, 0.9)$
and $(\tau, b_{\rm out})=(0.56r/{\rm r_{d}}+2.71\,{\rm Gyr}, 0.9)$
with $68\%$ confidence level. The notation of the observed data
is the same as that of Fig. \ref{Fig:results1}. It can be seen from
Fig. \ref{Fig:results1_1} that the solid lines can nicely
reproduce all the observational radial profiles, which suggests that
this parameter group may reasonably describe the crucial
ingredients of the main physical processes that regulate the formation
and evolution of NGC\,300.

\subsection{Stellar disk growth}

\begin{figure}
  \centering
  \includegraphics[angle=0,scale=0.58]{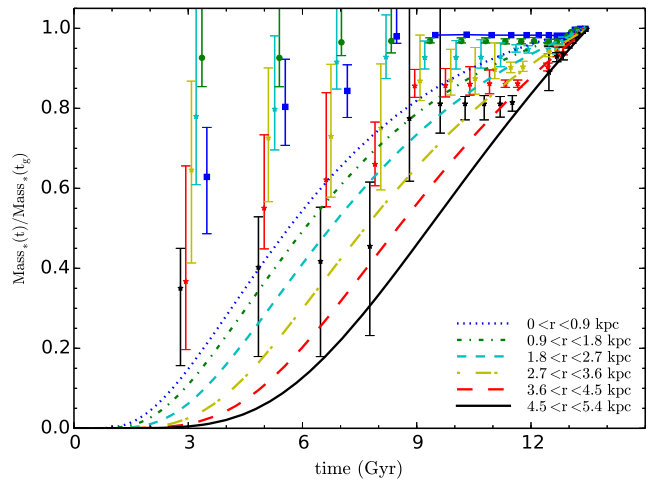}
    \caption{Relative stellar mass growth of six spatial
    components, including 0-0.9\,kpc (dotted lines), 0.9-1.8\,kpc
    (dot-dashed lines), 1.8-2.7\,kpc (dashed lines), 2.7-3.6\,kpc
    (long dot-dashed lines), 3.6-4.5\,kpc (long dashed lines) and
    4.5-5.4\,kpc (solid lines).The stellar masses in the different components
    are normalized to their corresponding present-day values. The observations
    are from Fig.6 of \citet{gogarten10}, and the data have been offset from
    one another to avoid overlapping error bars.
    }
  \label{Fig:cumulative}
\end{figure}

   \begin{figure}
  \centering
  \includegraphics[angle=0,scale=0.58]{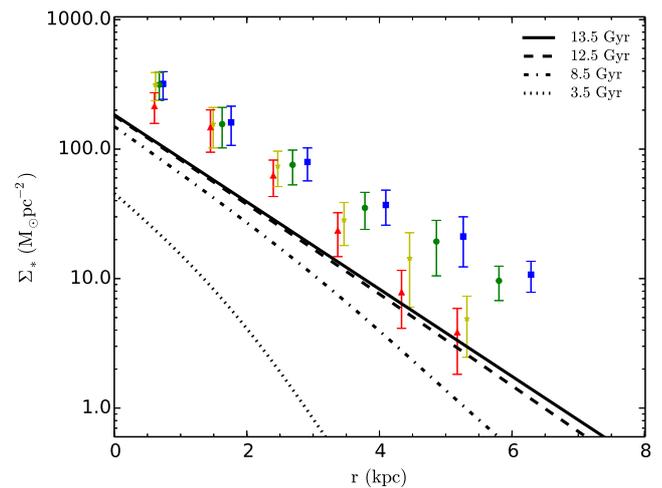}
    \caption{Time evolution of the stellar mass surface
    density profiles for the disk of NGC\,300. Different line types
    represent the radial profile of stellar mass surface density at
    3.5\,Gyr (dotted lines), 8.5\,Gyr(dot-dashed), 12.5\,Gyr(dashed
    lines) and 13.5\,Gyr(solid lines). The observations are from
    Fig.7 of \citet{gogarten10}, and the data have been offset from
    one another to avoid overlapping error bars.
    }
  \label{Fig:starfrofile}
\end{figure}

   \begin{figure}
  \centering
  \includegraphics[angle=0,scale=0.7]{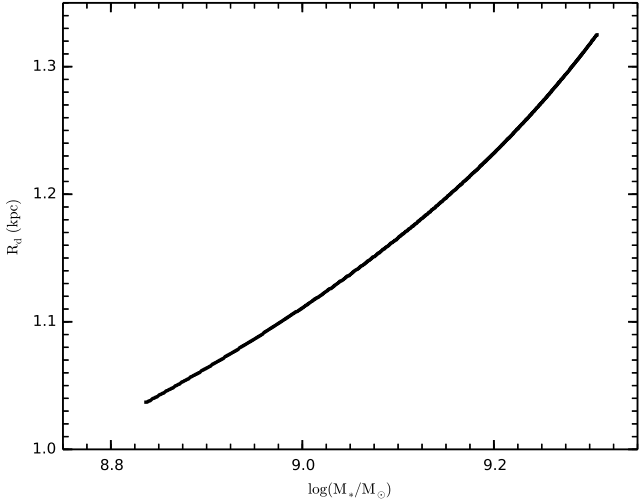}
    \caption{Time evolution of the stellar-mass-size (i.e.,
    the disk scale-length) for the
    disk of NGC\,300 from $z=1$ to $z=0$.
    }
  \label{Fig:M_S}
\end{figure}

In order to quantify the inside-out assembly of
NGC\,300, Fig. \ref{Fig:cumulative} plots the best-fitting predictions of the growth
curves of stellar masses in different regions, including 0-0.9\,kpc
(dotted lines), 0.9-1.8\,kpc (dot-dashed lines), 1.8-2.7\,kpc
(dashed lines), 2.7-3.6\,kpc (long dot-dashed lines),
3.6-4.5\,kpc (long dashed lines) and 4.5-5.4\,kpc (solid lines),
and compare them with the measurements from
Fig.6 of \citet{gogarten10}. Note that the stellar masses in Fig. \ref{Fig:cumulative}
are normalized by their present-day values. Fig. \ref{Fig:starfrofile} displays
the time evolution of the stellar mass surface density
profile at 3.5\,Gyr (dotted lines), 8.5\,Gyr(dot-dashed),
12.5\,Gyr(dashed lines) and 13.5\,Gyr(solid lines).

It can be seen from Fig. \ref{Fig:cumulative}
that the stellar mass in the corresponding regions has been
steadily increasing to its present-day value and the outer
parts of the disk formed a greater fraction of their stars
at recent times than the inner parts of the disk, which is consistent
with the "inside-out" formation scenario \citep{kim04,MM07,gogarten10}.
Moreover, Fig. \ref{Fig:starfrofile} shows that the stellar mass surface density
has changed substantially in the outer regions of the disk than in the
inner regions, in line with the trend of Fig.7 in \citet{gogarten10}.
However, the best-fitting model predictions fail to
reproduce the measurements of Fig.6 and Fig.7 from \citet{gogarten10}.
The possible reasons may be
identified as a) the stellar mass of NGC\,300 in our work is
$\sim 1.928\times10^{9}\rm M_{\odot}$ derived from $K-$band
luminosity \citep{MM07}, but it in \citet{gogarten10} is about
$\sim 7.0\times10^{9}\rm M_{\odot}$;
b) the galactocentric distances are within 5.4\,kpc for
the disk of NGC\,300 in \citet{gogarten10},
while ours are extend to 8.0\,kpc;
or a combination of these two. Thus, it is difficult to directly
compare our results with the measurements from \citet[][Fig. 6 and Fig. 7]{gogarten10}.
Most importantly, the main trend of our best-fitting model
predictions are in accordance with the information of Fig. 6 and Fig. 7
in \citet{gogarten10}.

Fig. \ref{Fig:M_S} also shows the stellar-mass-size
evolution (from $z=1$ to $z=0$) for the disk of NGC\,300. It can
be found that both the stellar mass and the scale-length of
NGC\,300 have been growing since $z=1$,
which is consistent with the results of the previous studies
\citep{brooks11,brook12}.

\begin{figure*}
   \centering
   \includegraphics[angle=0,scale=0.9]{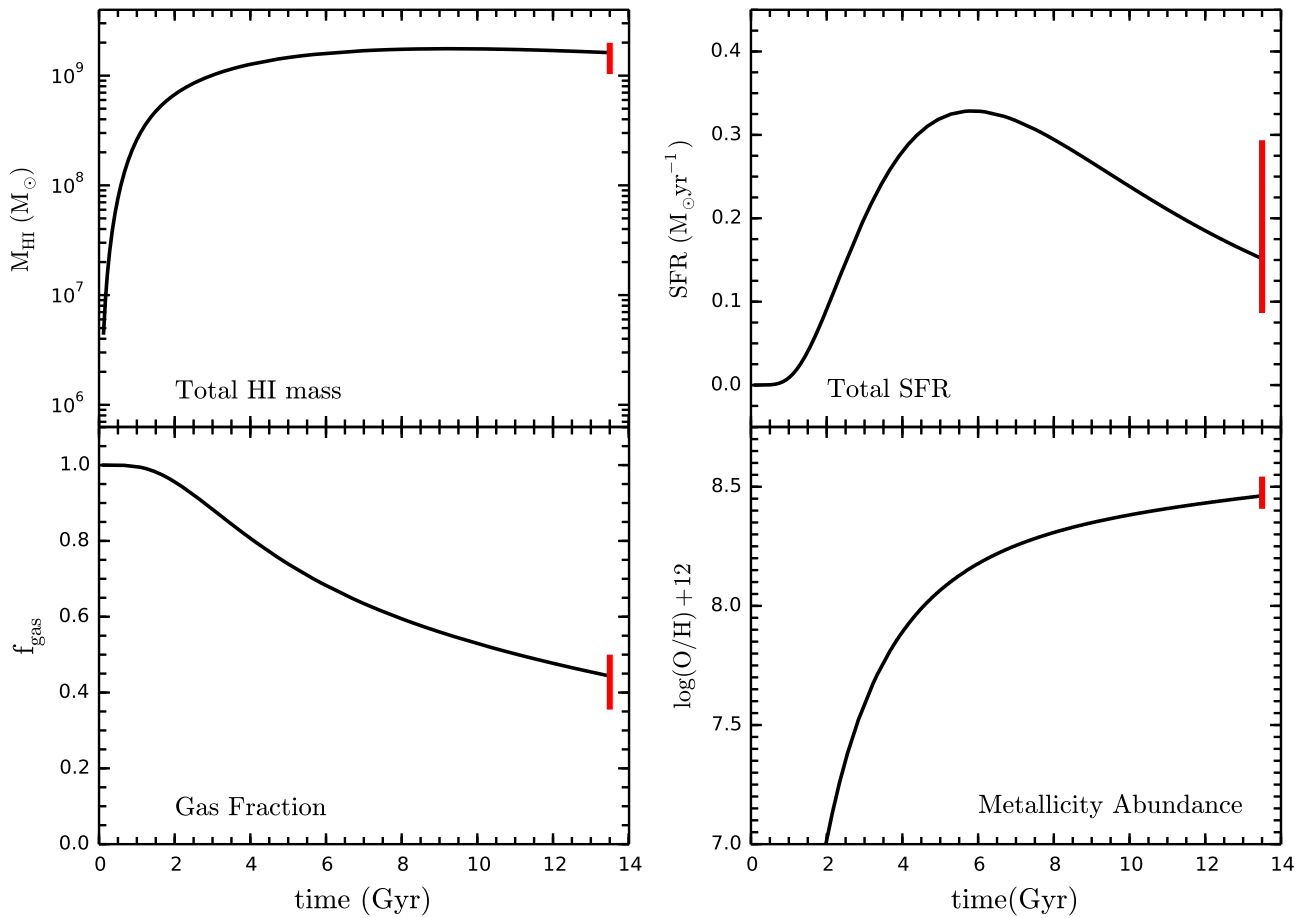}
   \caption{Time evolution of global atomic hydrogen
   gas mass, SFR, characteristic oxygen abundance, and the total gas
   fraction of NGC\,300 disk. Vertical bars at 13.5\,Gyr
   represent observational constraints (see Table \ref{Tab:obs2}).}
   \label{Fig:results2}
\end{figure*}

\subsection{Global evolution}

The evolution of the total amount of atomic hydrogen gas, SFR,
characteristic $\rm 12+log(O/H)$ (defined as the oxygen value
$\rm 12+log(O/H)$ at the effective radius $r_{\rm eff}$, which
is equal to 1.685 times the radial scalelength $r_{\rm d}$ of
the disk), and of the gas fraction
$f_{\rm gas}$ predicted by the best-fitting model are shown by
solid lines in Fig. \ref{Fig:results2}. They are compared to the
observational data displayed in Table \ref{Tab:obs2},
which are plotted as red vertical bars at the present time, i.e.,
$t\,=\,13.5\,\rm Gyr$.

Fig. \ref{Fig:results2} shows that the best-fitting model
predictions reproduce fairly well with the global observational
constraints of the NGC\,300 disk. It can be also found that, in
the first $\sim1$\,Gyr  the atomic hydrogen gas assembles very
fast, but the gas fraction does not decrease and there is no star
formation occurring. Although it is generally believed that
almost all star formation occurs in molecular clouds, the cold gas
in the earlier stage is mainly in the form of atomic gas.
After the amount of atomic hydrogen exceeds a critical value,
the atomic hydrogen gas begins to convert into molecular hydrogen
gas and then the star formation process accelarates. Fig. \ref{Fig:results2}
also shows that the oxygen abundance increases during the
whole evolution history,
and there is greater abundance increase before the evolution age
$t\sim7\,\rm Gyr$ compared to the later epoch.
This is mainly due to the fact that, in the later epoch, as the infall
rate of cold gas gradually decreases with time, so too does the
accumulation of molecular hydrogen gas, thus the SFR decreased
as well and the oxygen abundance enrichment becomes relatively slow.

\subsection{Comparison with M33}

\begin{table}
\caption[best]{The main ingredients and parameters of the best-fitting models for NGC\,300 and M33.}
\begin{center}
\begin{tabular}{lll}
\hline
\hline
General           &     Prescription       &     Parameter                \\
\hline
Age of disk (Gyr)                           &     13.5                          \\
IMF                                         &     \citet{Chabrier03}                    \\
Mass limits ($\rm M_{\odot}$)               &     0.1--100                    \\
Yield                                       &     $(X_{i})_{\odot}$             \\
SFR ($\rm M_{\odot}pc^{-2}Gyr^{-1}$)        &     $\Sigma_{\rm H_{2}}(r,t)/t_{\rm dep}$   &  $t_{\rm dep}$  \\
Infall rate ($\rm M_{\odot}pc^{-2}Gyr^{-1}$) &    $\propto t\cdot e^{-t/\tau}$      &  $\tau$ \\
Outflow rate ($\rm M_{\odot}pc^{-2}Gyr^{-1}$) &   $b_{\rm out}\Psi(r,t)$  &  $b_{\rm out}$ \\
Metallicity of infall gas                  &      $Z_{f_{\rm in}}=0$                     \\
$12+\rm {log(O/H)_{\odot}}$                &      $8.69\,^{\rm a}$                \\
\hline
Individual         &     NGC\,300          &     M33           \\
\hline
Total stellar mass ($10^{9}\rm M_{\odot}$)  &    1.928      & 4.0                 \\
Scale-length $r_{\rm d}$ (kpc)              &    1.29       & 1.4                   \\
$t_{\rm dep} (\rm Gyr)$                     &    1.9        &   0.46     \\
Infall time-scale $\tau(r)$(Gyr)            &  $0.35r/{r_{\rm d}}+2.47$   &  $r/{r_{\rm d}}+5.0$  \\
Outflow efficiency $b_{\rm out}$            &    0.9        &  $0.5$         \\
\hline
\end{tabular}
\end{center}
Note:\\
$^{\rm a}$ The solar oxygen abundance value is from \citet{asplund09}.
\label{tab:best}
\end{table}

M33 is nearly a twin to NGC\,300 in Hubble Type and mass, and we
have already explored the evolution and SFH of M33 in a previous
work \citep{kang12}. In this section, we will compare the SFH of
NGC\,300 predicted by the best-fitting model with that of M33.
The main ingredients and parameters of the best-fitting models
for NGC\,300 and M33 are presented
in Table \ref{tab:best}, and the reader is referred to \citet{kang12}
for a more indepth description of the physical details for the model
of M33.
\begin{figure*}
   \centering
   \includegraphics[angle=0,scale=0.58]{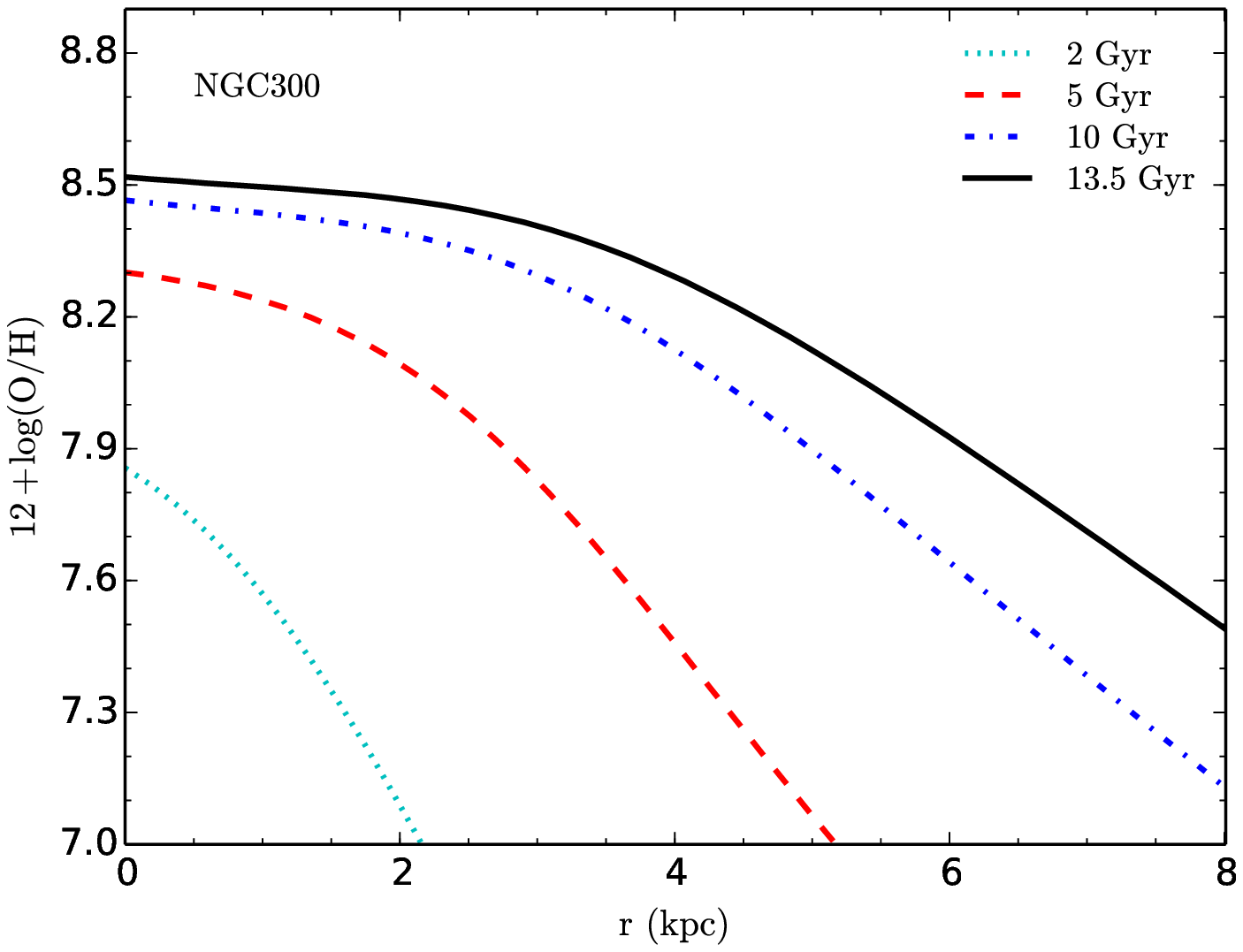}
   \includegraphics[angle=0,scale=0.58]{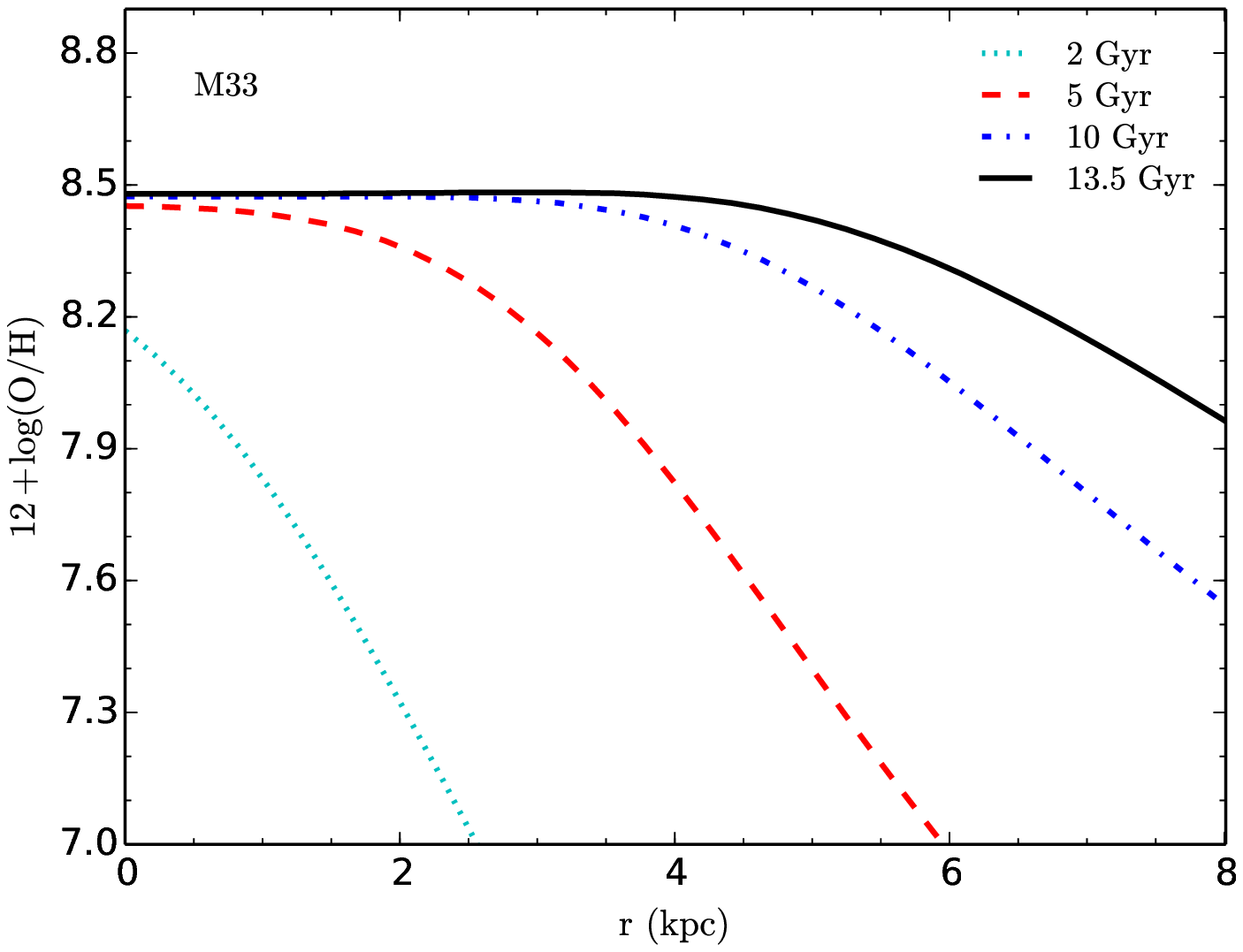}
   \caption{ The time evolution of the radial profiles of $\rm 12+log(O/H)$ for
   the disks of NGC\,300 (left panel) and M33 (right panel) predicted by their
   own best-fitting models. Different line types represent the radial profile
   of $\rm 12+log(O/H)$ at 2 (dotted line),
   5 (dashed line), 10 (dot-dashed line), and 13.5\,Gyr (solid line).
   }
   \label{Fig:Z_profile}
\end{figure*}


First of all, we plot the time evolution of the gas-phase oxygen
abundance gradient of NGC\,300 (left panel) and that of M33
(right panel) in Fig. \ref{Fig:Z_profile}. Different
line types represent the radial profile of $\rm 12+log(O/H)$ at
different time, i.e., 2 (dotted line), 5 (dashed line), 10 (dot-dashed line), and at 13.5\,Gyr
(solid line). The best-fitting model predicted present-day
(i.e., $t\,=\,13.5\,\rm Gyr$) abundance gradients
are $-$0.0843$\rm\,dex\,kpc^{-1}$ for NGC\,300 disk
and $-0.0219\rm\,dex\,kpc^{-1}$ for M33 disk from 0.2 to 6\,kpc in radius,
which is in good agreement with the observed gradient derived in
the same radial range, i.e., $-$0.0842$\pm$0.0065 for NGC\,300 from
\citet{pilyugin14_1} and $-$0.027$\pm$0.012 for M33 from \citet{RS08}.
It can be seen from Fig. \ref{Fig:Z_profile} that NGC\,300 has
a steep radial metallicity gradient and a significant increase in
metallicity with time. As for M33,
it has a flatter gradient and a slower increase in
metallicity with time, especially in the inner parts of the disk,
which indicates that M33 may experience more gas infall at later times
to dilute the metallicity. Furthermore,
we find that the metallicity in NGC\,300 has increased
with time in all radial bins, suggesting that there is less
infall of primordial gas recently. This is generally in accordance with
the observed finding of \citet{gogarten10}.

\begin{figure}
  \centering
  \includegraphics[angle=0,scale=0.58]{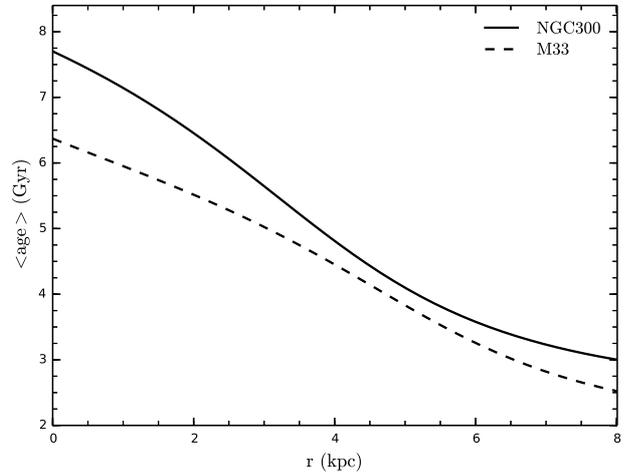}
    \caption{Current radial profiles of mean stellar age for NGC\,300 (solid line)
    and M33 (dashed line) predicted by their own best-fitting models.
    }
  \label{Fig:age}
\end{figure}

In order to demonstrate clearly the property of the stellar age
of both NGC\,300 and M33, we plot the mean stellar age along the
disks of them predicted by their own best-fitting models with the
solid line for NGC\,300 and with the dashed line for M33 in
Fig. \ref{Fig:age}. It can be seen that there exists radial age
gradient in both NGC\,300 and M33's disks, that is, the cumulative
age distribution shifts to younger ages as the radius increases,
which is consistent with the inside-out growth of the stellar disk
in NGC\,300 and M33 \citep{kim04,MM07,Williams09,gogarten10}.
This is mainly due to the fact that
the short infall time-scale means that a large fraction of stars
formed at the early stage and hence high mean stellar age, and vice
verse. Fortunately, \citet{gogarten10}
studied the cumulative SFH for the disk of NGC\,300 out to 5.4\,kpc,
and they found that $>\,90\%$ of the stars are older than 6\,Gyr
in the inner regions, while only $\sim\,40\%$ of stars are this
old in the outermost parts (i.e., $4.5\,<r<\,5.4\,\rm kpc$).
Meanwhile, the model predicted mean age of stellar populations along the
NGC\,300 disk is 7.70\,Gyr in the central region, 3.85\,Gyr at
$r\,=\,5.4\,\rm kpc$ and 2.97\,Gyr at $r\,=\,8.0\,\rm kpc$,
respectively. As for M33, \citet{barker11} concluded that the mean
age of stellar population at $r\,=\,9.1\,\rm kpc$ in M33 is $\sim2-4\,\rm Gyr$,
and our model predicted age in this place is 2.33\,Gyr.
Furthermore, our model predicted mean stellar age in the central
region of M33 is $\sim6.37\,\rm Gyr$. \citet{Williams09} fitted the
stellar disk growth within the inner disk
of M33, and found that most of the stars
formed from 10\,Gyr to 5\,Gyr ago. All of these suggest that
our model predictions are basically in agreement with the previous
observational results \citep{Williams09,gogarten10,barker11}.
What's more, it can be also found that the stellar age of
NGC\,300 is older than that of M33 throughout the studied region,
especially for the inner region of the NGC\,300 disk, and the mean
stellar age difference between the inner parts of the two galaxy
disks is up to $\sim1.33\,\rm Gyr$, while that between the outmost
regions of them is $\sim0.49\,\rm Gyr$. The older age of stars
in NGC\,300 than that in M33 indicates a lack of infall
of primordial gas and a small fraction of stars formed at later
times in NGC\,300 disk.

\begin{figure*}
   \includegraphics[angle=0,scale=0.585]{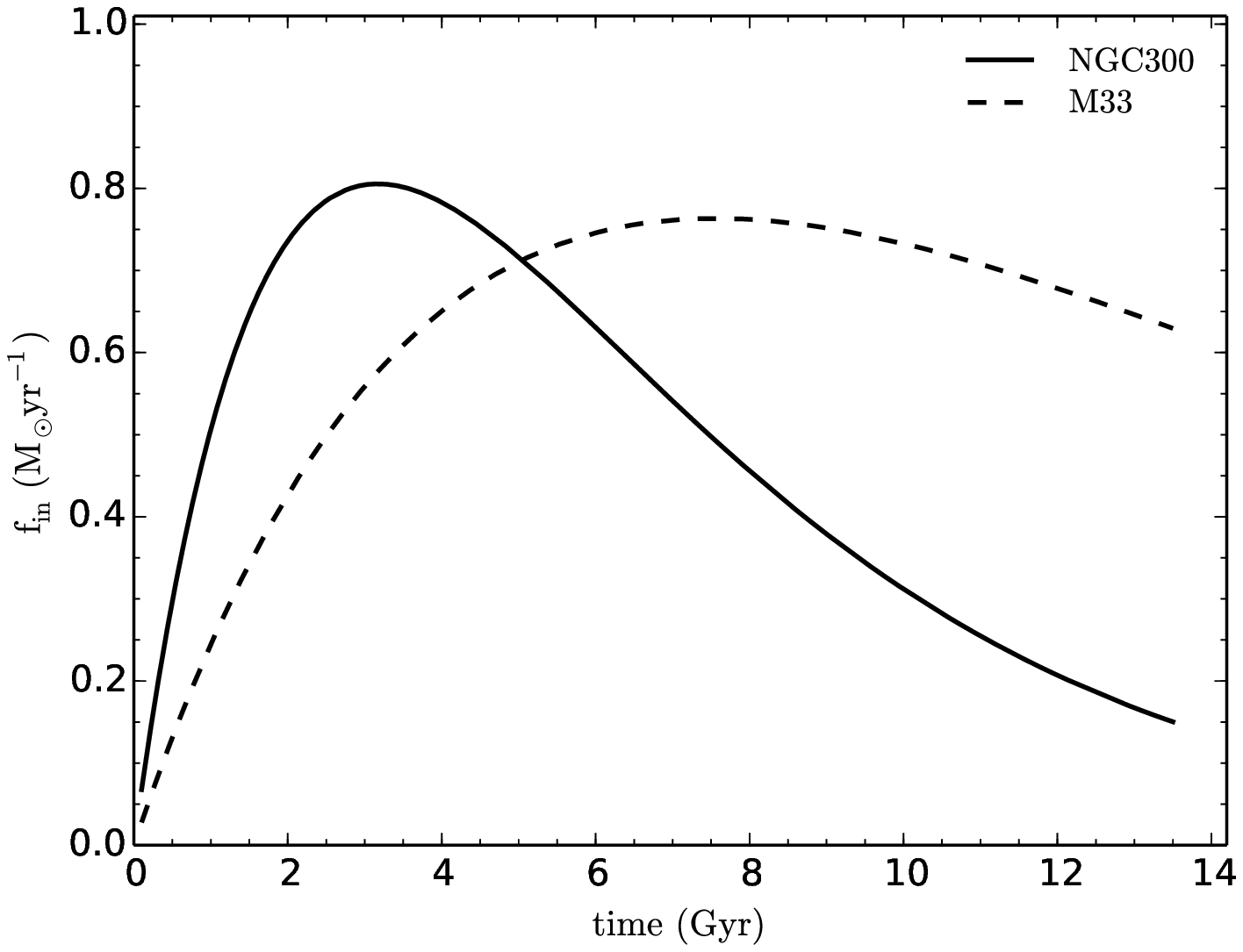}
   \includegraphics[angle=0,scale=0.585]{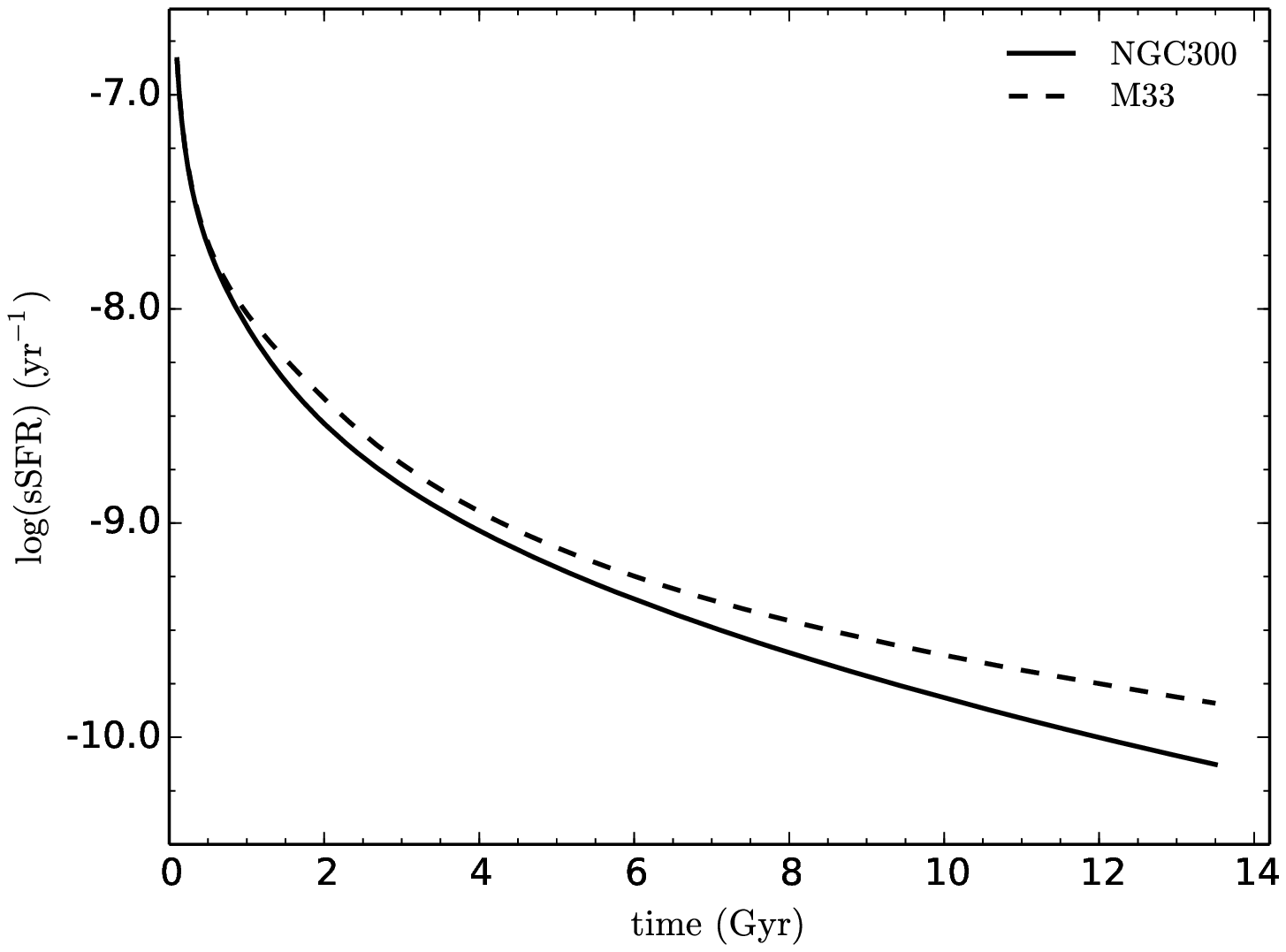}
   \caption{The evolution of gas infall rate (left panel) and sSFR (right
   panel) for NGC\,300 (solid line)
   and M33 (dashed line) predicted by their own best-fitting models. }
\label{Fig:gas_infall}
\end{figure*}

Specific star formation rate (sSFR) is defined as the ratio
of the current SFR to the current stellar mass,
$\rm sSFR\,=\,SFR/M_{\ast}$. Thus, higher values of the sSFR indicate
that a larger fraction of stars were formed recently, and it can be
used to characterize the SFH of galaxies.
The time evolution of gas infall rate and sSFR are shown in the
left and right panel of Fig. \ref{Fig:gas_infall}, respectively.
It can be seen from the left panel of Fig. \ref{Fig:gas_infall}
that the infall rate of NGC\,300 given by our best-fitting model is
gradually increasing with time and reaches its peak at
about 10\,Gyr ago, and then slowly drops down, while that of M33
achieves its peak around 6\,Gyr ago and then gradually falls.
Moreover, the infall rate of NGC\,300 increases and falls faster
than that of M33, and the present-day infall rate of NGC\,300
is much lower than that of M33. This indicates that the buildup
of the stellar disk of NGC\,300 is faster than that of M33.
At the same time, the right
panel of Fig. \ref{Fig:gas_infall} reveals that the sSFR of NGC\,300
is lower than that of M33 during the whole evolution history,
and the difference between them increases with time and
reaches the maximum at the present time. This is also
a suggestive of less gas infall and less active in star formation
recently along the disk of NGC\,300 than that of M33.

\begin{figure}
  \centering
  \includegraphics[angle=0,scale=0.55]{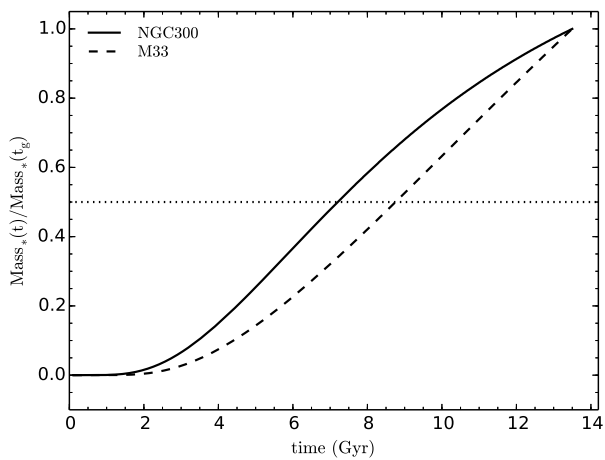}
    \caption{Stellar mass growth histories of NGC\,300 (solid line) and M33
    (dashed line) predicted by their own best-fitting models. Stellar masses
    are normalized to their present-day values, and the horizontal
    dotted line in the panel marks when the stellar mass achieves 50\% of its
    final value.
    }
  \label{Fig:stellar}
\end{figure}

In Fig. \ref{Fig:stellar}, we display the best-fitting model predictions
of the evolution of the stellar mass for both NGC\,300 (solid line)
and M33 (dashed line). To make the growth history of $M_{\ast}$ more
visible, stellar masses are normalized to their present-day values,
and to compute the $M_{\ast}$ growth rate, the horizontal dotted
line in the panel marks when the stellar mass achieves 50\% of its
final value. It can be seen that both NGC\,300 and M33 have
been steadily increasing to their present-day values. What's
more, half of the total stellar mass of NGC\,300 has been
assembled during the last $\sim6$\,Gyr, while that of M33
has been accumulated during the last $\sim4.5$\,Gyr. That is
to say, the mean age of stellar population in NGC\,300 is
older than that in M33.

\begin{figure}
  \centering
  \includegraphics[angle=0,scale=0.65]{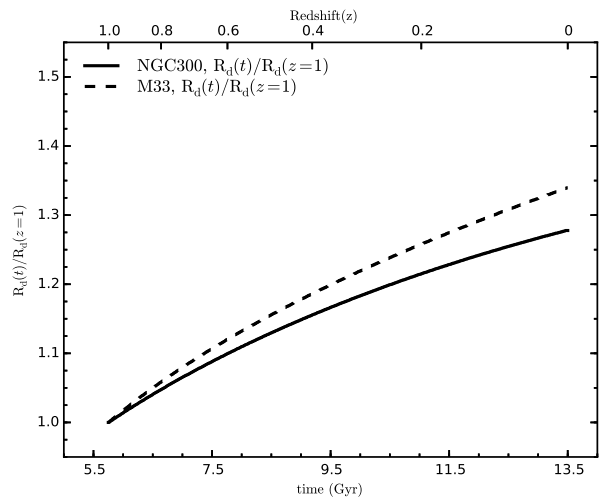}
    \caption{Growth of the disk scale-lengths for both
    NGC\,300 (solid line) and M33 (dashed line) predicted by their
    own best-fitting models. The disk scale-lengths are normalized to
    their sizes at $z=1$.
    }
  \label{Fig:MS}
\end{figure}

Fig. \ref{Fig:MS} plots the evolution of disk scale-length
of NGC\,300 (solid line) and M33 (dashed line) predicted by their
own best-fitting models, which are normalized to their values at $z=1$.
It can be seen that
the disk sizes of both NGC\,300 and M33 show clear
growth. Moreover, the scale-length of M33 changes more significant
than that of NGC\,300, which indicates that there is less fraction
of stars formed recently in the disk of NGC\,300 than that of M33.

To sum up, the aforementioned results suggest that the
stellar population of NGC\,300 is older than that of M33.
There is a lack of primordial gas infall onto the disk
of NGC\,300 and a less fraction of stars formed recently
in NGC\,300 than in M33.
Our results reinforce the recent results of \citet{gogarten10}
and \citet{Guglielmo15} that despite the two galaxies with
similar stellar mass and morphology, they have experienced
different histories due to their environmental differences.

\section{Summary}
\label{sect:summary}

NGC\,300 is a bulge-less and isolated low-mass disk galaxy and has not
experienced migration during its evolution history. In this work,
we build a bridge for the disk galaxy NGC\,300 between its observed
properties and its evolution history by constructing a simple
chemical evolution model. Some of our conclusions are as follows.

\begin{enumerate}

\item Our results show that the model predictions are very
sensitive to the adopted infall time-scale, but
the outflow process plays an important role in
shaping the gas-phase metallicity profiles along the disk of
NGC\,300 since it takes a large fraction of metals away from its disk
during its evolution history.

\item Using the classical $\chi^{2}$ methodology,
we compare the model results with the corresponding observations to
find the best combination of $a$, $b$ and $b_{\rm out}$ (i.e., 0.35,
2.47 and 0.9). As a result,
the model adopting an inside-out formation scenario
(i.e., $\tau\,=\,0.35r/{\rm r_{d}}+2.47\,{\rm Gyr}$) and an
appropriate gas outflow rate (i.e., $b_{\rm{out}}=0.9$) can
successfully reproduce the observed constraints of NGC\,300.
Our results suggest that NGC\,300 likely experiences a rapid
growth of its disk and the stellar population of it is
predominately old, consistent with the previous observed
results \citep{bulter04,Tikhonov05,gogarten10}.

\item We also compared the best-fitting model predicted evolution history of
NGC\,300 with that of M33. We find that the mean stellar age of
NGC\,300 is older than that of M33, and there is a lack of primordial
gas infall onto the disk of NGC\,300 and a less fraction of stars
formed recently in NGC\,300 than in M33. The comparative study of
the two bulge-less systems shows that
local environmental difference may paly an important part in the
secular evolution of the galaxy disks.

\end{enumerate}
\begin{acknowledgements}
We thank the anonymous referee whose comments and suggestions
have improved the quality of this paper greatly.
Xiaoyu Kang and Fenghui Zhang are supported by the National
Natural Science Foundation of China (NSFC)
grant No. 11403092, 11273053, 11033008, 11373063, and Yunnan
Foundation No. 2011CI053.
Ruixiang Chang is supported by the NSFC grant No.11373053, 11390373,
and Strategic Priority Research Program "The Emergence of
Cosmological Structures" of the Chinese Academy of Sciences
(CAS; grant XDB09010100).
\end{acknowledgements}

\bibliographystyle{aa}
\bibliography{kxy}


\end{document}